\documentclass[runningheads]{llncs}
\usepackage[T1]{fontenc}
\usepackage{graphicx}
\usepackage{caption}
\usepackage[table]{xcolor}
\usepackage{array}
\usepackage{graphicx} 
\usepackage{float}
\usepackage{amsmath}
\usepackage{hyperref}
\usepackage{url}

\usepackage[final]{microtype}

\begin{document}

\title{RATNUS: Rapid, Automatic Thalamic Nuclei Segmentation using Multimodal MRI inputs}

\author{
Anqi Feng\inst{1, 2} \and
Zhangxing Bian \inst{1} \and
Blake E. Dewey \inst{3} \and
Alexa Gail Colinco \inst{4} \and\\
Jiachen Zhuo \inst{4} \and
Jerry L. Prince \inst{1}
}
\institute{
Department of Electrical and Computer Engineering, Johns Hopkins University, Baltimore, MD 21218, USA\\[0.4em] \and
Laboratory of Behavioral Neuroscience, National Institute on Aging, National~Institutes~of~Health, Baltimore, MD 21224, USA\\[0.4em] \and
Department of Neurology, Johns Hopkins School of Medicine, Baltimore,~MD,~21287,~USA\\[0.4em] \and
Department of Diagnostic Radiology and Nuclear Medicine,
University~of~Maryland~School~of~Medicine, Baltimore, MD 21218, USA
}

\authorrunning{A. Feng et al.}
\titlerunning{RATNUS}

\maketitle  
\begin{abstract}
Accurate segmentation of thalamic nuclei is important for better understanding brain function and improving disease treatment. 
Traditional segmentation methods often rely on a single T1-weighted image, which has limited contrast in the thalamus. 
In this work, we introduce RATNUS, which uses synthetic T1-weighted images with many inversion times along with diffusion-derived features to enhance the visibility of nuclei within the thalamus. Using these features, a convolutional neural network is used to segment 13 thalamic nuclei. 
For comparison with other methods, we introduce a unified nuclei labeling scheme. Our results demonstrate an 87.19$\%$ average true positive rate~(TPR) against manual labeling. In comparison, FreeSurfer and THOMAS achieve TPRs of 64.25\% and 57.64\%, respectively, demonstrating the superiority of RATNUS in thalamic nuclei segmentation. The source code and pretrained models are available at: \url{https://github.com/ANQIFENG/RATNUS}.
\end{abstract}

\section{Introduction}
The thalamus is a vital brain structure located between the cortex and brain stem, composed of multiple functional clusters called nuclei~\cite{fama2015thalamic,sherman2001exploring,van2003contributions}. These nuclei have significant relevance to diseases such as Parkinson's diseases, Alzheimer's disease, multiple sclerosis, and frontotemporal dementia~\cite{henderson2000loss,braak1991alzheimer,minagar2013thalamus,bocchetta2020thalamic}, making their  segmentation crucial for targeted research and the development of therapies. Despite this, many studies typically segment the thalamus as one entity in part because the exclusive use of standard T1-weighted and T2-weighted images lack the contrast necessary to distinguish the small nuclei within the thalamus~\cite{billot2023synthseg,huo20193d,wachinger2018deepnat,woolrich2001temporal}. 

Existing methods that provide segmentations of certain thalamic nuclei include FreeSurfer~\cite{iglesias2018probabilistic} and THOMAS~\cite{su2019thalamus}. FreeSurfer uses a Bayesian segmentation approach with a probabilistic atlas of the human thalamus as prior knowledge; however, its use  of only the Magnetization Prepared Rapid Gradient Echo~(MPRAGE) sequence~\cite{mugler1990three} limits the contrast between nuclei. THOMAS uses multi-atlas segmentation and label fusion techniques on high-contrast, white-matter-nulled MPRAGE scans, also known as Fast Gray Matter Acquisition T1 Inversion Recovery~(FGATIR)~\cite{sudhyadhom2009high}, collected at 7T to segment 12 thalamic nuclei. While it achieves good results on 7T images, its effectiveness on 3T scans has not been quantitatively validated. None of these methods use multiple T1-weighted images with different inversion times~(multi-TI), which provide higher thalamus contrast not achievable by MPRAGE or FGATIR alone.

Diffusion Magnetic resonance imaging~(MRI)~\cite{battistella2017robust,stough2013isbi,stough2014automatic,tregidgo2023accurate,tregidgo2023domain,wiegell2003automatic,yan2023spie,ziyan2006segmentation} has been used for thalamic nuclei segmentation. One approach \cite{stough2014automatic} develops a hierarchical random forest framework to parcellate the thalamus into 6 nuclei, utilizing inputs such as fractional anisotropy~(FA), the Knutsson 5D vector and edge map, and connectivity with cortical regions. However, its manual labeling was based on FA and Knutsson edge maps, which does not align perfectly with labels derived from T1-weighted images. Two approaches~\cite{tregidgo2023accurate,tregidgo2023domain} extend the previous FreeSurfer method by incorporating the FA and principal eigenvector with MPRAGE images. The first approach~\cite{tregidgo2023accurate} uses Bayesian segmentation but lacks a model for partial volume effects, leading to poor performance
at resolutions worse than 1mm. The second approach~\cite{tregidgo2023domain}, based on a convolutional neural network~(CNN), has not been quantitatively compared with ground truth for its segmentation of 23 histological labels. When evaluating its performance on a 10-label grouping, Dice scores were below 0.6, indicating limitations in their segmentation accuracy.

In this paper, we present a new approach to thalamic nuclei segmentation, named RATNUS. Its innovation lies in the integration of multi-TI images and features derived from diffusion images, providing enhanced contrast for accurate thalamic nuclei segmentation. 
Our contributions are threefold:
(1)~We present a method to correctly synthesize multi-TI images from MPRAGE and FGATIR images.
(2)~We develop a CNN-based network to identify 13 distinct thalamic nuclei, trained using novel manual delineations.
(3)~We establish a protocol for unifying different labeling schemes across methods to facilitate comparison. RATNUS achieves superior performance compared to benchmark methods, demonstrating its outstanding capability in the segmentation of thalamic nuclei.

\section{Methods}
\subsection{Data and Manual Labels}
Our dataset is derived from a study on mild traumatic brain injury~(MTBI), which was approved by the University of Maryland's ethics review board. The datasest contains 24 subjects, including 14 healthy controls and 10 people with MTBI. Each participant underwent MRI, including MPRAGE, FGATIR, and diffusion sequences. The MPRAGE and FGATIR images were acquired using the same sequence protocol: an isotropic resolution of 1 mm, repetition time~(TR): 4000 ms, echo time~(TE): 3.37 ms, and inversion time~(TI): MPRAGE 1400 ms and FGATIR 400 ms. 
Diffusion images were acquired with an isotropic resolution of 2mm, with 136 directions, at b-values of 0, 1000, and 2500, using an anterior-posterior~(AP) phase-encoding direction with additional posterior-anterior~(PA) b0 images for distortion correction. All data were collected using a 3T Siemens Prisma scanner. 

\begin{figure}[!tb]
\centering
\includegraphics[width = 1\textwidth]{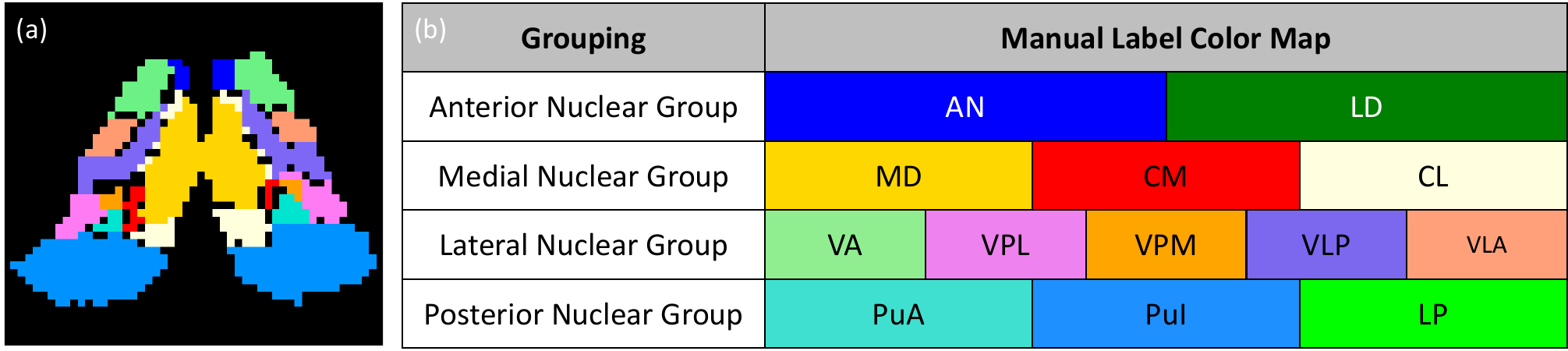}
\caption{(a)~Our manual labels of thalamic nuclei are sparse.
(b)~Color-coded labeling scheme for thalamic nuclei according to the Morel atlas.}
\label{fig:color_coded_labeling_scheme}
\end{figure}

Manual thalamic nuclei delineation was guided by the Morel Atlas~\cite{morel2007stereotactic}. Referencing both the T$_{1}$ map and multi-TI images, labels were assigned only to those voxels for which the rater had high confidence of correct label assignment. This led to sparse labels, ensuring high confidence in the labeled voxels but leaving many voxels unlabeled, particularly in boundaries between nuclei, as illustrated in Fig.~\ref{fig:color_coded_labeling_scheme}(a). Our study focuses on 13 nuclei (and nuclear groups): Anterior Nucleus~(AN), Central Lateral~(CL), Center Median~(CM), Lateral Dorsal~(LD), Lateral Posterior~(LP), Mediodorsal~(MD), Anterior Pulvinar~(PuA), Inferior Pulvinar~(PuI), Ventral Anterior~(VA), Ventral Lateral Anterior~(VLA), Ventral Lateral Posterior~(VLP), Ventral Posterior Lateral~(VPL), and Ventral Posterior Medial~(VPM). The labeling scheme of these thalamic nuclei with color coding is shown in Fig.~\ref{fig:color_coded_labeling_scheme}(b).

\subsection{Data Processing Pipeline}
We developed a pipeline to synthesize multi-TI images. We also generated comprehensive features from the diffusion images. These processed images and features greatly enhance the visibility of details within the thalamus and are concatenated together to form the input to the models.

\vspace{-4 mm}
\subsubsection{Structural MRI Processing}
Our processing of MPRAGE and FGATIR images begins with a coarse-to-fine registration, using ANTS \cite{avants2009advanced} for rigid registration specific to each subject. Initially, MPRAGE images are registered to the MNI space. This is followed by the registration of FGATIR images to their corresponding MPRAGE images, transforming both images to the same atlas space.

Following registration, the MPRAGE and FGATIR images go through two processing steps: N4 Bias Field Correction~\cite{tustison2010n4itk} and White Matter Mean Normalization~\cite{reinhold2019evaluating}. To accurately derive proton density~(PD) and T$_{1}$ maps from the MPRAGE and FGATIR images, as described below, it is important to process these images together, as separate adjustments in brightness or contrast will introduce errors in computation of the PD and T$_{1}$ maps~\cite{tohidi2023joint}. Consequently, we use a harmonic bias field computed as the geometric mean of the two N4 bias fields, found separately from the MPRAGE and FGATIR images, for uniform intensity normalization. We also create a white matter mask from the MPRAGE image using Fuzzy C-means clustering~\cite{reinhold2019evaluating} and then adjust the two images with the same scaling factor, determined by the average white matter intensity within the MPRAGE image.

\begin{figure}[!tb]
\centering
\includegraphics[width = 1\textwidth]{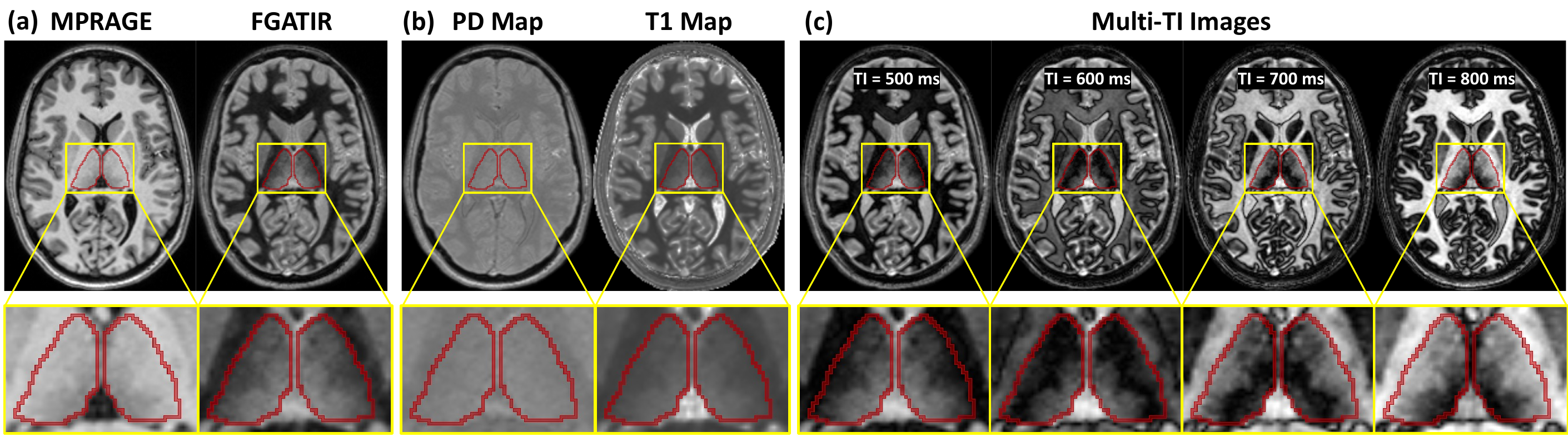}
\caption{Different T1-weighted images. (a)~MPRAGE and FGATIR images after registration, N4 bias field correction and white matter mean normalization. (b)~PD and T$_{1}$ maps derived from MPRAGE and FGATIR images. (c)~Example multi-TI images at TIs: 500 ms, 600 ms, 700 ms, and 800 ms. Thalamus regions are highlighted with yellow boxes and delineated with red contours.}
\label{fig:data}
\end{figure}

For a given voxel within a T1-weighted image, the signal intensity can be expressed as
\begin{equation}
I^{(i)} = \mathrm{PD}^{(i)} [ 1 - 2e^{(-\mathrm{TI} / T_1^{(i)})} + e^{(-\mathrm{TR} / T_1^{(i)})} ], 
\label{formula 1}
\end{equation}
where $I^{(i)}$ denotes the image intensity for the $i$-th voxel, $\mathrm{PD}^{(i)}$ the proton density map value for the $i$-th voxel, $T_1^{(i)}$ the $T_1$ map value for the $i$-th voxel, TI the inversion time, and TR the repetition time.
As the MPRAGE and FGATIR images were acquired in identical fashion except for their different TIs, we are able to write two distinct equations based on Equation~\ref{formula 1} for each image and then solve for the PD and T$_1$ maps using nonlinear least squares.
We then proceed to synthesize a series of multi-TI images by keeping PD and T$_1$ values constant and systematically varying the TI value. 
Specifically, we varied TI from 400 to 1400~ms in increments of 20~ms, resulting in a set of 51 images, revealing contrast between different nuclei in different images. 
Fig.~\ref{fig:data} shows examples of MPRAGE and FGATIR images after processing, corresponding PD and T$_{1}$ maps, and some computed multi-TI images.

\subsubsection{Diffusion MRI Processing}
We processed our diffusion MRI data using the Tortoise software \cite{pierpaoli2010tortoise}, starting with the DIFFPREP module for denoising, motion, and distortion corrections, followed by DR-BUDDI module for EPI distortion correction leveraging opposite phase encoding directions. These two steps also registered the diffusion data to their corresponding T2-weighted images in MNI space. We then used the DIFFCALC module to perform tensor fitting and calculate key metrics, including axial diffusivity~(AD), fractional anisotropy~(FA), radial diffusivity~(RD), Trace, Eigenvalues, and Eigenvectors, along with the three Westin measures~(WS, WP, WL).
Following this, we applied the Knutsson mapping to convert the principal eigenvector into a 5-dimensional~(5D) vector~\cite{knutsson1985capaidm,stough2014automatic}.
Moreover, we generated an edge map based on these orientations, serving as a visual representation of the changes in direction of the principal eigenvectors. Using the Knutsson 5D vector and the Knutsson edge map helps to better define the boundaries and reveal more subtle details within the thalamus, surpassing the conventional eigenvector depictions. 

\subsection{Thalamic Nuclei Segmentation}
\begin{figure}[!tb]
\centering
\includegraphics[width = 0.98\textwidth]{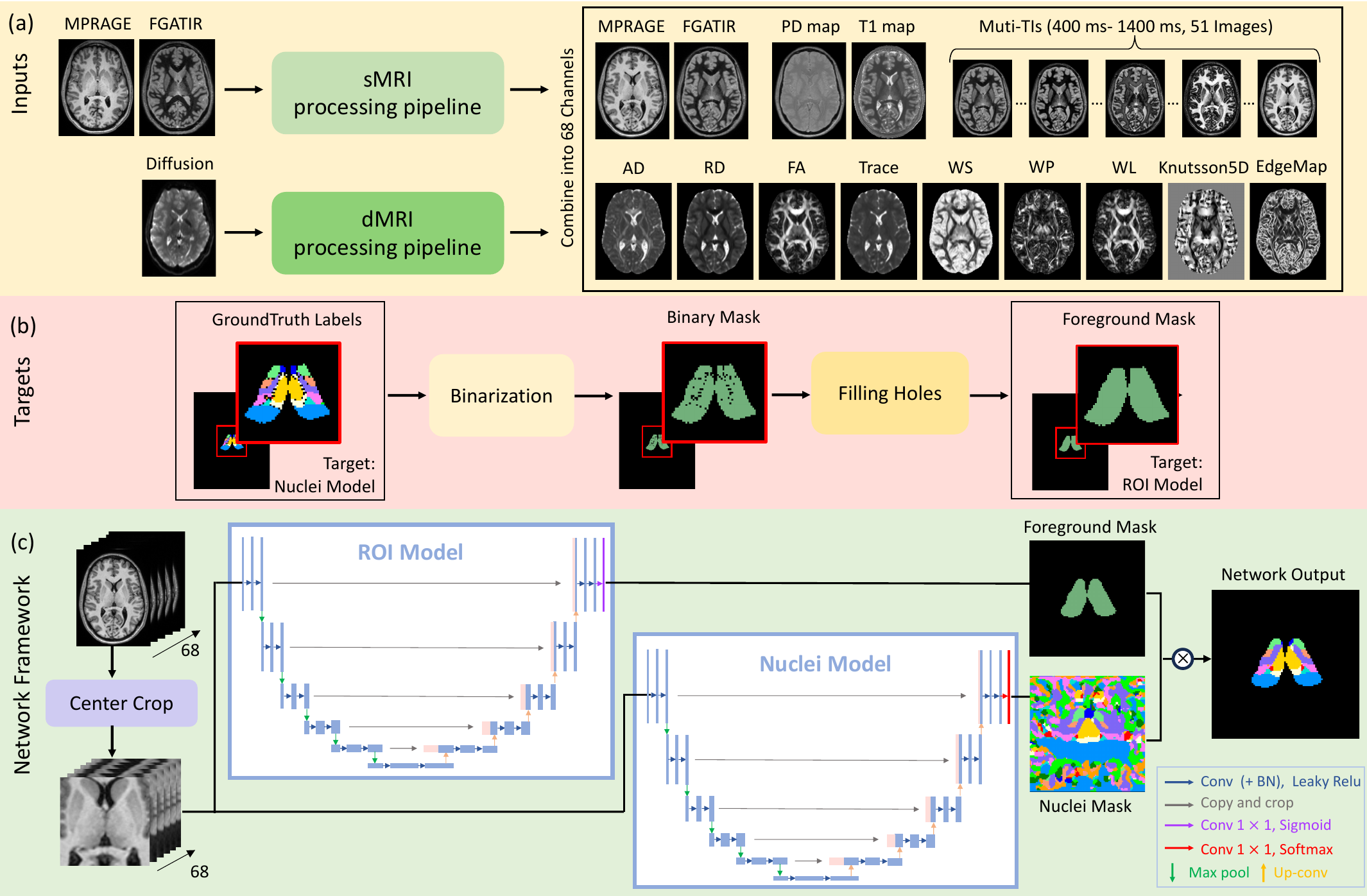}
\caption{Overview of the thalamic nuclei segmentation framework. (a)~The processing for the structural~(sMRI) and diffusion MRI~(dMRI), leading to a combined 68-channel network input. (b)~The generation of the foreground mask (ground truth of the ROI model) using the thalamic nuclei labels (ground truth of the Nuclei model). (c)~The network framework, consists of the ROI model and the Nuclei model, both built upon a 5-level 3D U-Net. The final segmentation is produced by multiplying the foreground mask from the ROI model with the nuclei mask from the Nuclei model.}
\label{fig:framework}
\end{figure}

Our work focuses on segmenting 13 distinct nuclei (and nuclear groups) within the thalamus. We implemented a two-step approach, with two networks each built upon a 5-level 3D U-Net~\cite{cciccek20163d}.
The first network identifies the region of interest~(ROI) and the second delineates the ROI into 13 nuclei classes, as illustrated in Fig.~\ref{fig:framework}.
For both models, the input consists of a 68-channel image per subject, derived from the processing pipeline described in Section~2.2, as shown in Fig.~\ref{fig:framework}(a). 
Volumes are center-cropped to $96\times 96\times 96$ voxels (specified in MNI coordinates), which always includes the entire thalamus. 

The first model, the ROI model, generates foreground masks that identify the areas of interest for the thalamic nuclei we aim to segment.
Before training, we convert the nuclei labels into a binary format to create a binary ground truth.
Given our sparse manual labels, we used the paint tool in the 3D Slicer Segmentation Editor to fill the holes within the binary ground truth without altering its overall external shape, as depicted in Fig.~\ref{fig:framework}(b).
The second model, the Nuclei model, segments the entire data into 13 classes, with each voxel being classified into one of the thalamic nuclei.
The last step is to multiply the foreground mask from the ROI model with the nuclei mask from the Nuclei model to generate the 13 thalamic nuclei predictions.

\subsubsection{Loss}
We trained the ROI model using Dice Loss \cite{milletari2016v}. When training the Nuclei model, we aim to minimize the Dice Loss for the 13 nuclei classes.
However, given the sparse ground truth, where the rater only labeled voxels with high confidence, leaving many voxels in the thalamus unlabeled, we applied a masking technique during the Dice Loss calculation. Specifically, we multiplied both the numerator and denominator of the Dice Loss by a ground truth mask (where labeled voxels are 1 and unlabeled voxels are 0) to ensure that only labeled voxels contributed to the loss calculation, thereby excluding the influence of unlabeled voxels. This naturally leads to the maximization of the true positive rate. The loss function for the Nuclei model is then formed as:

\begin{equation}
\mathcal{L}_{\mathrm{Nuclei}} = 1 - \frac{1}{13} \sum_{c=1}^{13} \frac{\sum_{i} T_{c,i} P_{c,i} + \varepsilon}{\sum_{i} T_{c,i} + \varepsilon}, 
\label{formula:loss2}
\end{equation}
where $T_{c,i}$ denotes the true label for class $c$ at voxel $i$, $P_{c,i}$ is the predicted probability for class $c$ at voxel $i$, and $\varepsilon$ is a smoothing factor set to 1e-6. 

\subsubsection{Implementation Details}
We performed an 8-fold cross-validation on 24 subjects with a training, validation, testing split ratio of 19:2:3, ensuring that all subjects were included in the testing process once. Both models are 3D U-Nets with 5 levels, trained using the Adam optimizer with a weight decay of 1e-4, and an initial learning rates of 1e-3, reduced by 20$\%$ if the validation loss does not improve over 5 epochs. We employed Leaky ReLU activation and Batch Normalization across all convolutional blocks. The random seed was set to 1234. The ROI model was trained using a sigmoid activation in the final layer, with a batch size of 4 and stopped training if the validation loss does not decrease within 10 epochs. The Nuclei model was trained using a softmax activation in the final layer, with a batch size of 1, and the same stopping criterion as the ROI model.
 
\subsection{Labeling Schemes Unifying Protocol}
\begin{figure}[H]
\centering
    \includegraphics[width = 0.77\textwidth]{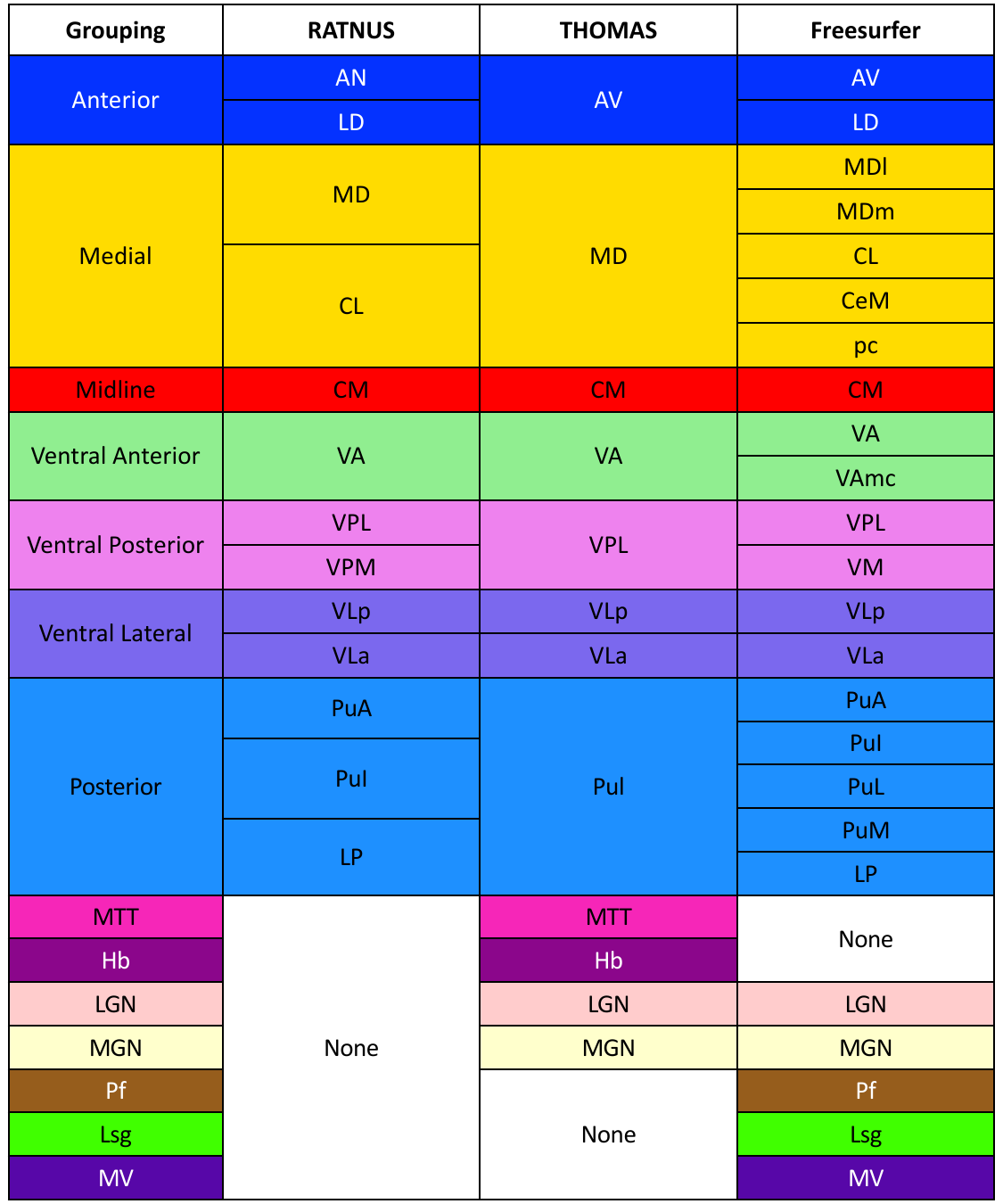}
    \caption{An exhaustive mapping of thalamic nuclei classes between RATNUS, and those defined by THOMAS and FreeSurfer. The first column lists the major categories derived from our grouping strategy, each of which corresponds to the merging of several nuclei in the second, third and, forth column. All major categories and individual instances are assigned a unique color for clear differentiation. Abbreviations for THOMAS and FreeSurfer nuclei are carefully extracted from~\cite{su2019thalamus} and~\cite{iglesias2018probabilistic}.}
\label{fig:alignment_of_thalamic_nuclei}
\end{figure}

We compared RATNUS to FreeSurfer~\cite{iglesias2018probabilistic} and THOMAS~\cite{su2019thalamus}, two popular approaches for thalamic nuclei segmentation. Since each method uses a unique set of rules for labeling thalamic nuclei, we used the Morel atlas \cite{morel2007stereotactic} as a structural guide to propose a unified labeling scheme for comparison. After reviewing correspondences between the schemes, we formed the following seven main categories: Anterior group, Medial group, Midline group, Ventral Anterior group, Ventral Posterior group, Ventral Lateral group, and Posterior group.
Nuclei that do not have a corresponding match---e.g., the MTT is labeled in the THOMAS atlas but is not represented in either our atlas or the FreeSurfer atlas, and the Lsg is labeled in FreeSurfer but is missing from ours and the THOMAS atlas---were marked as distinct classifications. The grouping and color-coding scheme is shown in Fig.~\ref{fig:alignment_of_thalamic_nuclei}.

\vspace{-3mm}

\section{Experiments and Results}
In this section, we evaluate the performance of the RATNUS model. We first compare RATNUS' segmentation results with ground truth labels to measure accuracy. Then, we benchmark RATNUS against FreeSurfer\cite{iglesias2018probabilistic} and THOMAS~\cite{su2019thalamus} to highlight its advantages.
As our ground truth labels are sparse, where the rater only labeled voxels with high confidence, leaving many voxels in the thalamus unlabeled, using Dice Score as the evaluation metric could be inappropriate, since it may count predictions as incorrect if they match the nuclei class but lack corresponding ground truth labels.
Therefore, we adopt the TPR as the evaluation metric for the following experiments. Additionally, as mentioned in the implementation details, we performed an 8-fold cross-validation on 24 subjects. This ensures that each subject is included in the testing process once, and the results presented below are the average and standard deviation calculated across all 24 subjects.

\subsection{Comparison Against Ground Truth}
\begin{figure}[!tb]
\centering
\includegraphics[width = 0.8\textwidth]{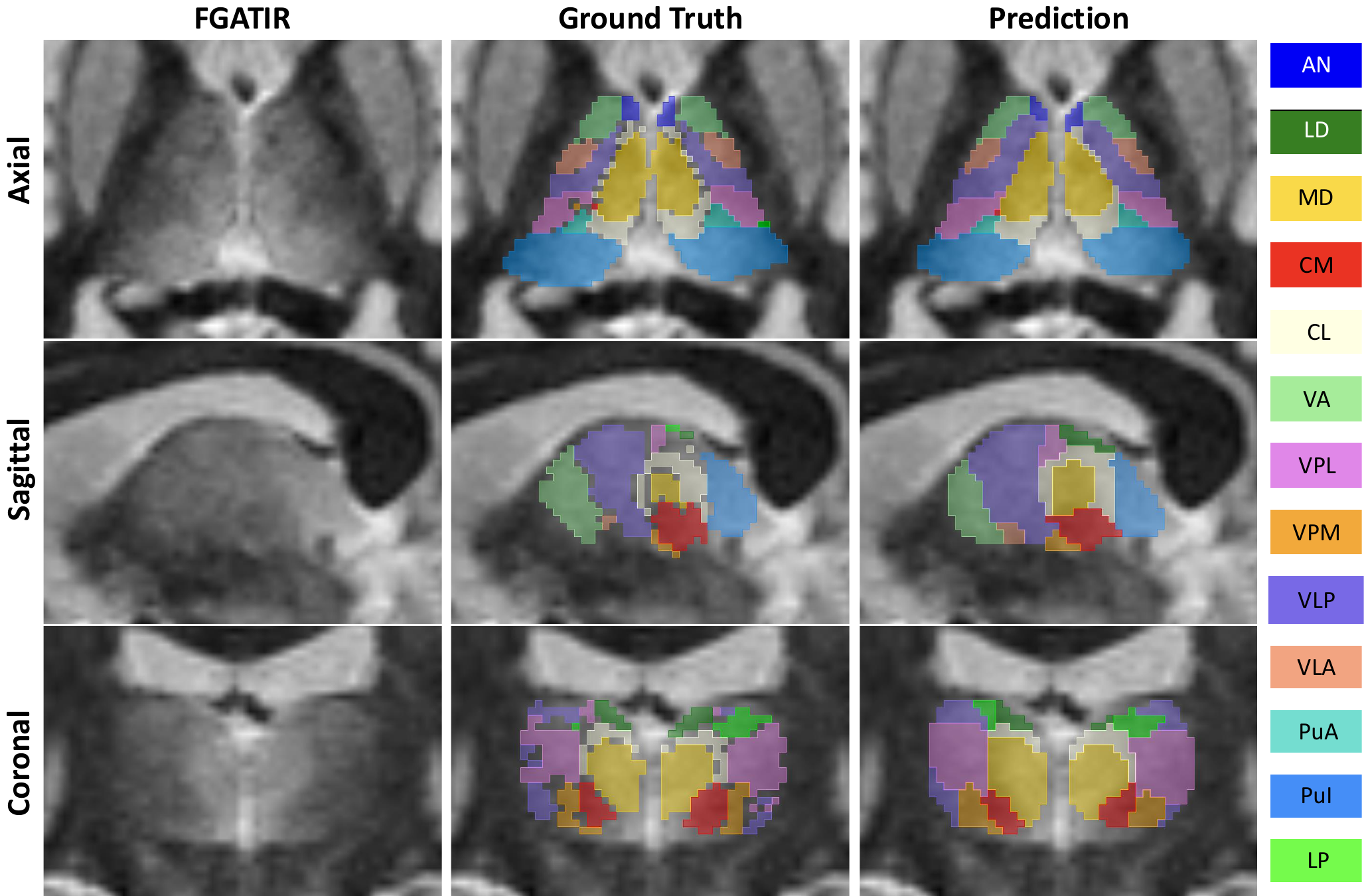}
\caption{Comparisons between model's predictions and ground truth labels. From left to right, each column shows the FGATIR scans, the ground truth, and our model predictions.  The provided color codes is consistent with Fig.~\ref{fig:color_coded_labeling_scheme}.}
\label{fig:8-fold_cross_validation_results}
\end{figure}

 We evaluated the RATNUS’s segmentation results by comparing them with ground truth labels using TPR. The quantitative results, summarized in Table~\ref{tab:average_TPR_scores}, illustrates RATNUS's consistently high TPR across all thalamic nuclei. Fig.~\ref{fig:8-fold_cross_validation_results}, provides a visual comparison, showing close alignments between our model's predictions and the ground truth, demonstrating RATNUS's capability in accurate thalamic nuclei segmentation.
 
\begin{table}[!tb]
\centering
\caption{TPRs for 13 thalamic nuclei, shown in percentages. ``Average'' indicates the volume-weighted mean across all nuclei.}
\label{tab:average_TPR_scores}
\resizebox{0.95\textwidth}{!}{
\begin{tabular}{|c|c|c|c|c|c|c|c|c|c|c|c|c|c|}
\hline
AN & CM & LD & LP & MD & PuA & PuI & VA & VLa & VLp & VPL & VPM & CL & Average \\ \hline
86.32 & 80.65 & 80.75 & 85.30 & 91.08 & 61.91 & 94.36 & 91.79 & 79.43 & 88.10 & 83.16 & 75.60 & 68.43 & 87.19 \\
$\pm$ & $\pm$ & $\pm$ & $\pm$ & $\pm$ & $\pm$ & $\pm$ & $\pm$ & $\pm$ & $\pm$ & $\pm$ & $\pm$ & $\pm$ & $\pm$ \\
1.58 & 5.10 & 4.06 & 3.53 & 1.31 & 7.82 & 0.92 & 2.21 & 4.77 & 1.93 & 3.64 & 7.98 & 2.53 & 1.47 \\ \hline
\end{tabular}}
\end{table}

\subsection{Comparisons Against FreeSurfer and THOMAS Segmentation}
\begin{figure}[!tb]
\centering
\includegraphics[width = 1\textwidth]{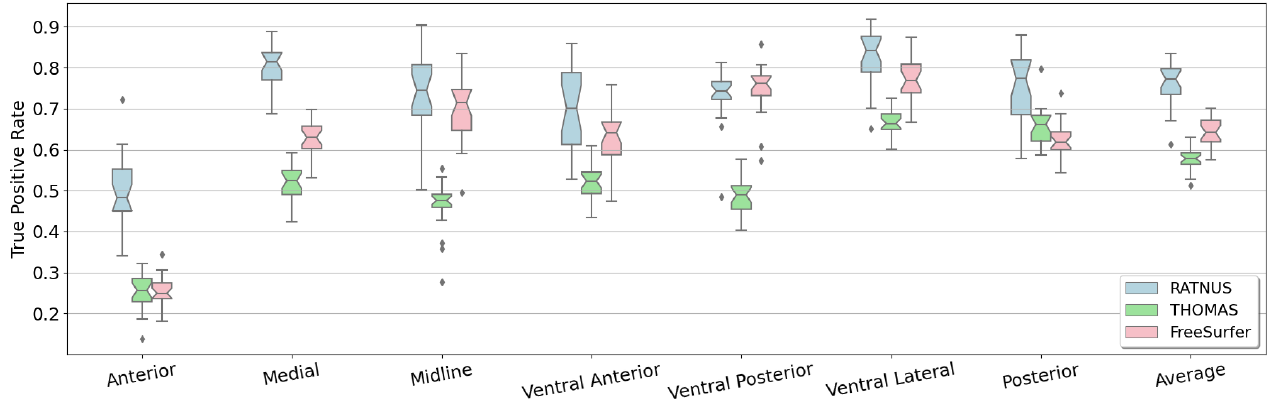}
\caption{Boxplot showing RATNUS's TPR compared with FreeSurfer and THOMAS. The horizontal axis lists major thalamic groups from Anterior to Posterior, with ``Average'' showing the volume-weighted mean of all the groups.}
\label{fig:boxplot_comparisions_pred_THOMAS_freesurfer}
\end{figure}

\begin{figure}[!tb]
\centering
    \includegraphics[width = 1\textwidth]{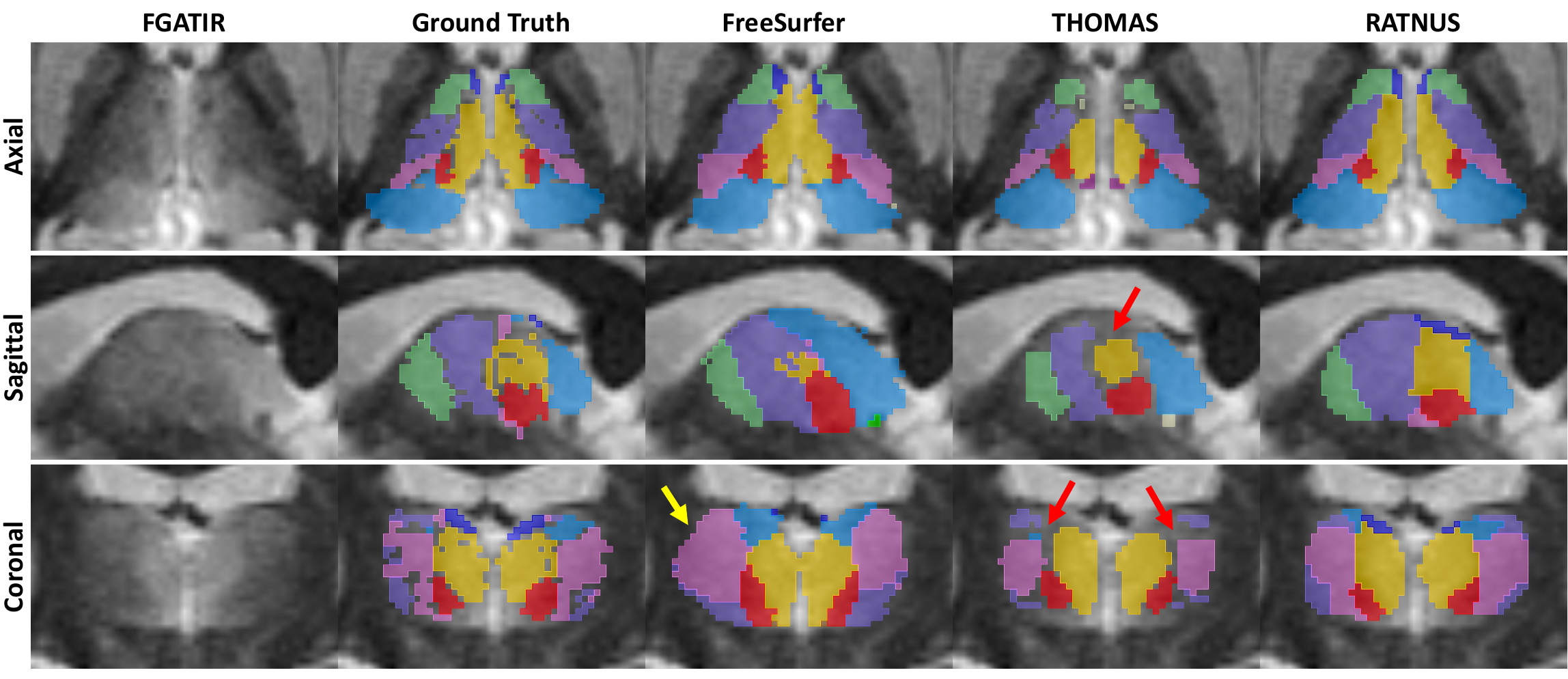}
    \caption{Comparative visualization of our model predictions against FreeSurfer and THOMAS across seven major thalamic regions. Unique instances, such as MGN, are also included. Color codes for each thalamic group are provided in Fig.~\ref{fig:alignment_of_thalamic_nuclei}.}
    \label{fig:Grouping_results}
\end{figure}

We compared our model's predictions with those of FreeSurfer and THOMAS after unifying them to a common ground using the protocol described in Section~2.4. The quantitative comparison focuses only on seven main groups and excludes unique labels and background. 
Fig.~\ref{fig:boxplot_comparisions_pred_THOMAS_freesurfer} shows that RATNUS consistently achieves higher TPR in all major thalamic groups, with an average TPR of 75.80\%, greatly surpassing THOMAS at 57.64\% and FreeSurfer at 64.25\%.
Qualitatively, as shown in Fig.~\ref{fig:Grouping_results}, RATNUS's predictions are more contiguous with fewer misclassified voxels and better adherence to the boundary of the thalamus, whereas FreeSurfer appears to over-segment (highlighted by the yellow arrow), and THOMAS tends to under-segment (highlighted by the red arrows) the thalamus.

\section{Conclusions and Discussions}
We propose RATNUS, a method that segments 13 thalamic nuclei within a CNN framework using the combination of multi-TI images and diffusion-derived features, which provide enhanced contrast within the thalamus. It achieves superior performance over FreeSurfer and THOMAS, demonstrating RATNUS's potential for precise thalamic nuclei segmentation.

Our method has some limitations. First, the approach benefits greatly from multimodal data, particularly multi-TI images, which provide detailed information about thalamic nuclei.
These multi-TI images require both MPRAGE and FGATIR sequences to generate. If either MPRAGE or FGATIR is unavailable, the multi-TI images cannot be produced, and the method cannot utilize the valuable information they provide.
Second, we compared our method to FreeSurfer and THOMAS, both of which use different labeling schemes. To ensure fairness, we proposed a unifying protocol based on the Morel atlas to standardize these schemes.
However, since our model was trained using our specific labels and tested on the same distribution, it is more likely to perform better than THOMAS and FreeSurfer, introducing some bias despite our unifying efforts.
Third, delineating thalamic nuclei is difficult and costly due to their small size and low contrast.
As a result, in this work, we only trained and tested models using 24 pairs of data and thalamic nuclei labels. 
Although this is comparable to the 28 delineations used to develop THOMAS, it is still somewhat limited for training deep learning models, potentially restricting model's generalization ability and robustness.

To address these limitations, our future work will focus on several key areas.
First, in order to eliminate the dependence on FGATIR, we plan to develop a network to synthesize multi-TI images from alternative multi-contrast imaging data, similar to~\cite{hays2024spie}.
Second, we plan to perform out-of-distribution tests to evaluate the generalization capabilities of FreeSurfer, THOMAS, and RATNUS.
Third, we plan to implement semi-supervised training techniques to leverage a larger amount of data which does not have ground truth labels.

\section*{Acknowledgments}
This work was supported in part by the Intramural Research Program of the National Institutes of Health~(NIH), National Institute on Aging.
It is also supported in part by the NIH through the National Institute of Neurological Disorders and Stroke grant R01-NS105503~(PIs: Zhuo \& Prince).

\bibliographystyle{splncs04}
\bibliography{reference}

\begin{thebibliography}{10}
\providecommand{\url}[1]{\texttt{#1}}
\providecommand{\urlprefix}{URL }
\providecommand{\doi}[1]{https://doi.org/#1}

\bibitem{avants2009advanced}
Avants, B.B., Tustison, N., Song, G.: Advanced normalization tools ({ANTS}). Insight j  \textbf{2}(365),  1--35 (2009)

\bibitem{battistella2017robust}
Battistella, G., et~al.: Robust thalamic nuclei segmentation method based on local diffusion magnetic resonance properties. Brain Structure and Function  \textbf{222},  2203--2216 (2017)

\bibitem{billot2023synthseg}
Billot, B., et~al.: {SynthSeg}: Segmentation of brain {MRI} scans of any contrast and resolution without retraining. Medical Image Analysis  \textbf{86},  102789 (2023)

\bibitem{bocchetta2020thalamic}
Bocchetta, M., et~al.: {Thalamic nuclei in frontotemporal dementia: Mediodorsal nucleus involvement is universal but pulvinar atrophy is unique to C9orf72}. Human Brain Mapping  \textbf{41}(4),  1006--1016 (2020)

\bibitem{braak1991alzheimer}
Braak, H., Braak, E.: Alzheimer's disease affects limbic nuclei of the thalamus. Acta neuropathologica  \textbf{81}(3),  261--268 (1991)

\bibitem{cciccek20163d}
{\c{C}}i{\c{c}}ek, {\"O}., et~al.: {3D U-Net}: learning dense volumetric segmentation from sparse annotation. In: Medical Image Computing and Computer-Assisted Intervention--MICCAI 2016: 19th International Conference, Athens, Greece, October 17-21, 2016, Proceedings, Part II 19. pp. 424--432. Springer (2016)

\bibitem{fama2015thalamic}
Fama, R., Sullivan, E.V.: Thalamic structures and associated cognitive functions: Relations with age and aging. Neuroscience \& Biobehavioral Reviews  \textbf{54},  29--37 (2015)

\bibitem{hays2024spie}
Hays, S.P., et~al.: {Revisiting registration-based image synthesis: A focus on unsupervised MR image synthesis}. In: Proceedings of SPIE Medical Imaging~(SPIE-MI 2024), San Diego, CA, February 18 -- 22, 2024. vol. 12926, pp. 257--265 (2024)

\bibitem{henderson2000loss}
Henderson, J., Carpenter, K., Cartwright, H., Halliday, G.: {Loss of thalamic intralaminar nuclei in progressive supranuclear palsy and Parkinson's disease: Clinical and therapeutic implications}. Brain  \textbf{123}(7),  1410--1421 (2000)

\bibitem{huo20193d}
Huo, Y., et~al.: {3D} whole brain segmentation using spatially localized atlas network tiles. NeuroImage  \textbf{194},  105--119 (2019)

\bibitem{iglesias2018probabilistic}
Iglesias, J.E., et~al.: A probabilistic atlas of the human thalamic nuclei combining ex vivo {MRI} and histology. NeuroImage  \textbf{183},  314--326 (2018)

\bibitem{knutsson1985capaidm}
Knutsson, H.: {Producing a Continuous and Distance Preserving 5-D Vector Representation of 3-D Orientation}. In: IEEE Computer Society Workshop on Computer Architecture for Pattern Analysis and Image Database Management. pp. 175--182 (1985)

\bibitem{milletari2016v}
Milletari, F., Navab, N., Ahmadi, S.A.: {V-Net}: Fully convolutional neural networks for volumetric medical image segmentation. In: 2016 Fourth International Conference on 3D Vision (3DV). pp. 565--571. IEEE (2016)

\bibitem{minagar2013thalamus}
Minagar, A., et~al.: The thalamus and multiple sclerosis: modern views on pathologic, imaging, and clinical aspects. Neurology  \textbf{80}(2),  210--219 (2013)

\bibitem{morel2007stereotactic}
Morel, A.: Stereotactic atlas of the human thalamus and basal ganglia. CRC Press (2007)

\bibitem{mugler1990three}
Mugler~III, J.P., Brookeman, J.R.: Three-dimensional magnetization-prepared rapid gradient-echo imaging ({3D MPRAGE}). Magnetic Resonance in Medicine  \textbf{15}(1),  152--157 (1990)

\bibitem{pierpaoli2010tortoise}
Pierpaoli, C., et~al.: {TORTOISE}: an integrated software package for processing of diffusion {MRI} data. In: ISMRM 18th Annual Meeting. vol.~1597. Stockholm (2010)

\bibitem{reinhold2019evaluating}
Reinhold, J.C., et~al.: Evaluating the impact of intensity normalization on {MR} image synthesis. In: Medical Imaging 2019: Image Processing. vol. 10949, pp. 890--898. SPIE (2019)

\bibitem{sherman2001exploring}
Sherman, S.M., Guillery, R.W.: Exploring the thalamus. Elsevier (2001)

\bibitem{stough2013isbi}
Stough, J.V., et~al.: {Thalamic Parcellation from Multi-Modal Data using Random Forest Learning}. In: 10$^{\mbox{\tiny{th}}}$ International Symposium on Biomedical Imaging~(ISBI~2013). pp. 852--855 (2013)

\bibitem{stough2014automatic}
Stough, J.V., et~al.: Automatic method for thalamus parcellation using multi-modal feature classification. In: Medical Image Computing and Computer-Assisted Intervention--MICCAI 2014. vol.~8675, pp. 169--176. Springer (2014)

\bibitem{su2019thalamus}
Su, J.H., et~al.: Thalamus optimized multi atlas segmentation ({THOMAS}): fast, fully automated segmentation of thalamic nuclei from structural {MRI}. NeuroImage  \textbf{194},  272--282 (2019)

\bibitem{sudhyadhom2009high}
Sudhyadhom, A., et~al.: A high re solution and high contrast {MRI} for differentiation of subcortical structures for {DBS} targeting: the {Fast Gray Matter Acquisition T1 Inversion Recovery} ({FGATIR}). NeuroImage  \textbf{47},  T44--T52 (2009)

\bibitem{tohidi2023joint}
Tohidi, P., et~al.: Joint synthesis of {WMn MPRAGE} and parameter maps using deep learning and an imaging equation. In: Medical Imaging 2023: Image Processing. vol. 12464, pp. 558--564. SPIE (2023)

\bibitem{tregidgo2023domain}
Tregidgo, H.F., Soskic, S., Olchanyi, M.D., et~al.: Domain-agnostic segmentation of thalamic nuclei from joint structural and diffusion {MRI}. arXiv preprint arXiv:2305.03413  (2023)

\bibitem{tregidgo2023accurate}
Tregidgo, H.F., et~al.: Accurate {Bayesian} segmentation of thalamic nuclei using diffusion {MRI} and an improved histological atlas. NeuroImage  \textbf{274},  120129 (2023)

\bibitem{tustison2010n4itk}
Tustison, N.J., et~al.: {N4ITK}: improved {N3} bias correction. IEEE Transactions on Medical Imaging  \textbf{29}(6),  1310--1320 (2010)

\bibitem{van2003contributions}
Van Der~Werf, Y.D., et~al.: Contributions of thalamic nuclei to declarative memory functioning. Cortex  \textbf{39}(4-5),  1047--1062 (2003)

\bibitem{wachinger2018deepnat}
Wachinger, C., Reuter, M., Klein, T.: {DeepNAT}: Deep convolutional neural network for segmenting neuroanatomy. NeuroImage  \textbf{170},  434--445 (2018)

\bibitem{wiegell2003automatic}
Wiegell, M.R., Tuch, D.S., Larsson, H.B., Wedeen, V.J.: Automatic segmentation of thalamic nuclei from diffusion tensor magnetic resonance imaging. NeuroImage  \textbf{19}(2),  391--401 (2003)

\bibitem{woolrich2001temporal}
Woolrich, M.W., et~al.: Temporal autocorrelation in univariate linear modeling of {FMRI} data. NeuroImage  \textbf{14}(6),  1370--1386 (2001)

\bibitem{yan2023spie}
Yan, C., et~al.: {Segmenting thalamic nuclei from manifold projections of multi-contrast MRI}. In: Proceedings of SPIE Medical Imaging~(SPIE-MI 2023), San Diego, CA, February 19 -- 23, 2023. vol. 12464, pp. 727--734 (2023)

\bibitem{ziyan2006segmentation}
Ziyan, U., Tuch, D., Westin, C.F.: Segmentation of thalamic nuclei from {DTI} using spectral clustering. In: Medical Image Computing and Computer-Assisted Intervention--MICCAI 2006: 9th International Conference, Copenhagen, Denmark, October 1-6, 2006. Proceedings, Part II 9. pp. 807--814. Springer (2006)

\end{thebibliography}

\end{document}